\documentclass[floatfix,a4paper,aps,pra,byrevtex,preprint,groupedaddress]{revtex4}
\usepackage{longtable}
\begin{document}
\newcommand{\et}{ and }
\newcommand{\oneauth}[2]{#1 #2,}
\newcommand{\erratum}[3]{\jcite{erratum}{#1}{#2}{#3}}
\newcommand{\inpress}[1]{{\it #1}}
\newcommand{\inbook}[5]{In {\it #1}; #2; #3: #4, #5}
\newcommand{\auth}[2]{#2, #1,}
\newcommand{\twoauth}[4]{#2, #1 and #4, #3,}
\newcommand{\jcite}[4]{#4, {\it #1} {\bf #2}, #3}
\newcommand{\andauth}[2]{and #2, #1,}
\newcommand{\book}[4]{#4, {\it #1} (#2, #3)}
\newcommand{\JCP}[3]{\jcite{J. Chem. Phys.}{#1}{#2}{#3}}
\newcommand{\JMS}[3]{\jcite{J. Mol. Spectrosc.}{#1}{#2}{#3}}
\newcommand{\JMSP}[3]{\jcite{J. Mol. Spectrosc.}{#1}{#2}{#3}}
\newcommand{\JMST}[3]{\jcite{J. Mol. Struct.}{#1}{#2}{#3}}
\newcommand{\JMSTR}[3]{\jcite{J. Mol. Struct.}{#1}{#2}{#3}}
\newcommand{\CPL}[3]{\jcite{Chem. Phys. Lett.}{#1}{#2}{#3}}
\newcommand{\CP}[3]{\jcite{Chem. Phys.}{#1}{#2}{#3}}
\newcommand{\PR}[3]{\jcite{Phys. Rev.}{#1}{#2}{#3}}
\newcommand{\JPC}[3]{\jcite{J. Phys. Chem.}{#1}{#2}{#3}}
\newcommand{\JPCA}[3]{\jcite{J. Phys. Chem. A}{#1}{#2}{#3}}
\newcommand{\JPCB}[3]{\jcite{J. Phys. Chem. B}{#1}{#2}{#3}}
\newcommand{\JPB}[3]{\jcite{J. Phys. B}{#1}{#2}{#3}}
\newcommand{\PRA}[3]{\jcite{Phys. Rev. A}{#1}{#2}{#3}}
\newcommand{\PRB}[3]{\jcite{Phys. Rev. B}{#1}{#2}{#3}}
\newcommand{\PRL}[3]{\jcite{Phys. Rev. Lett.}{#1}{#2}{#3}}
\newcommand{\JCC}[3]{\jcite{J. Comput. Chem.}{#1}{#2}{#3}}
\newcommand{\MP}[3]{\jcite{Mol. Phys.}{#1}{#2}{#3}}
\newcommand{\MPH}[3]{\jcite{Mol. Phys.}{#1}{#2}{#3}}
\newcommand{\APJ}[3]{\jcite{Astrophys. J.}{#1}{#2}{#3}}
\newcommand{\CPC}[3]{\jcite{Comput. Phys. Commun.}{#1}{#2}{#3}}
\newcommand{\CJC}[3]{\jcite{Can. J. Chem.}{#1}{#2}{#3}}
\newcommand{\CJP}[3]{\jcite{Can. J. Phys.}{#1}{#2}{#3}}
\newcommand{\JCSFII}[3]{\jcite{J. Chem. Soc. Faraday Trans. II}{#1}{#2}{#3}}
\newcommand{\BBPC}[3]{\jcite{Ber. Bunsenges. Phys. Chem.}{#1}{#2}{#3}}
\newcommand{\PCCP}[3]{\jcite{Phys. Chem. Chem. Phys.}{#1}{#2}{#3}} 
\newcommand{\FD}[3]{\jcite{Faraday Discuss.}{#1}{#2}{#3}}
\newcommand{\PRSA}[3]{\jcite{Proc. Royal Soc. A}{#1}{#2}{#3}}
\newcommand{\JACS}[3]{\jcite{J. Am. Chem. Soc.}{#1}{#2}{#3}}
\newcommand{\JOSA}[3]{\jcite{J. Opt. Soc. Am.}{#1}{#2}{#3}}
\newcommand{\IJQCS}[3]{\jcite{Int. J. Quantum Chem. Symp.}{#1}{#2}{#3}}
\newcommand{\IJQC}[3]{\jcite{Int. J. Quantum Chem.}{#1}{#2}{#3}}
\newcommand{\SAA}[3]{\jcite{Spectrochim. Acta A}{#1}{#2}{#3}}
\newcommand{\TCA}[3]{\jcite{Theor. Chem. Acc.}{#1}{#2}{#3}}
\newcommand{\TCAold}[3]{\jcite{Theor. Chim. Acta}{#1}{#2}{#3}}
\newcommand{\JPCRD}[3]{\jcite{J. Phys. Chem. Ref. Data}{#1}{#2}{#3}}
\newcommand{\SCIENCE}[3]{\jcite{Science}{#1}{#2}{#3}}
\newcommand{\CR}[3]{\jcite{Chem. Rev.}{#1}{#2}{#3}}
\newcommand{\ACIE}[3]{\jcite{Angew. Chem. Int. Ed.}{#1}{#2}{#3}}
\newcommand{\IJCK}[3]{\jcite{Int. J. Chem. Kinet.}{#1}{#2}{#3}}
\newcommand{\JCT}[3]{\jcite{J. Chem. Thermodyn.}{#1}{#2}{#3}}
\newcommand{\CEJ}[3]{\jcite{Chem. Eur. J.}{#1}{#2}{#3}}
\newcommand{\OM}[3]{\jcite{Organometallics}{#1}{#2}{#3}}
\newcommand{\NATURE}[3]{\jcite{Nature}{#1}{#2}{#3}}
\newcommand{\NSB}[3]{\jcite{Nature Struct. Bio.}{#1}{#2}{#3}}

\title{Alkali and Alkaline Earth Metal Compounds: Core-Valence Basis Sets and Importance
of Subvalence Correlation}

\author{Mark A. Iron, Mikhal Oren, and Jan M.L. Martin}

\affiliation{Department of Organic Chemistry, Weizmann Institute of Science, 76100
Re\d{h}ovot, Israel}

\email{comartin@wicc.weizmann.ac.il}

\homepage{http://theochem.weizmann.ac.il/}

\date{{\em Mol. Phys.}, in press (W. G. Richards special issue)}

\begin{abstract}
Core-valence basis sets for the alkali and alkaline earth metals 
Li, Be, Na, Mg, K, and Ca are proposed. The basis sets are validated
by calculating spectroscopic constants of a variety of diatomic 
molecules involving these elements. Neglect of $(3s,3p)$ correlation 
in K and Ca compounds will lead to erratic results at best, and
chemically nonsensical ones if chalcogens or halogens are present.
The addition of low-exponent $p$ functions to the K and Ca basis 
sets is essential for smooth convergence of molecular properties.
Inclusion of inner-shell correlation is important for accurate spectroscopic
constants and binding energies of all the compounds. In basis 
set extrapolation/convergence calculations, the explicit inclusion of
alkali and alkaline earth metal subvalence correlation at all steps is essential
for K and Ca, strongly recommended for Na, and
optional for Li and Mg, while in Be compounds, an
additive treatment in a separate `core correlation' step is probably sufficient.
Consideration of $(1s)$ inner-shell correlation energy in first-row elements 
requires inclusion of $(2s,2p)$ `deep core' correlation energy in K and Ca
for consistency. The latter requires special CCV$n$Z `deep core correlation' 
basis sets. For compounds involving Ca bound to electronegative elements, 
additional $d$ functions in the basis set are strongly recommended. For
optimal basis set convergence in such cases, we suggest the sequence
CV(D+$3d$)Z, CV(T+$2d$)Z, CV(Q+$d$)Z, and CV5Z on calcium.
\end{abstract}
\maketitle

\section{Introduction}

The rate-determining step in accurate wavefunction \emph{ab initio}
calculations is the determination of the correlation energy. Moderately
reliable methods like CCSD (coupled cluster with all single and double
substitutions\cite{Pur82}) have theoretical CPU time scalings $\propto n^{2}N^{4}$
, where $n$ and $N$ represent the numbers of electrons and basis functions,
respectively. For the CCSD(T) (i.e., CCSD with a quasiperturbative
estimate of the effect of connected triple substitutions\cite{Rag89})
method --- which has been termed the ``gold standard of quantum chemistry''
(T. H. Dunning) --- the corresponding scaling is $\propto n^{3}N^{4}$.

For this reason, the idea of only including the `chemically relevant'
valence electrons in the correlation problem and constraining the
inner-shell or `core' orbitals to be doubly occupied gained currency
early on. Bauschlicher et al.\cite{Bau88cc-cv} were the first to
propose a partitioning of the inner-shell correlation energy into a
core-core (CC) component --- involving excitations exclusively out
of the inner-shell orbitals --- and a core-valence (CV) component
--- involving simultaneous excitations out of the valence and inner-shell
orbitals. While, at least in elements with a large core-valence gap
(see below), the CC component can be expected to roughly cancel between
a molecule and its separated atoms, the CV component may affect chemically
relevant molecular properties to some degree.

Over time, it has become recognized that the inclusion of inner-shell
correlation --- even in first-row systems --- is important for accurate binding
energies (e.g. \cite{Mar95cc,G3cc}), molecular geometries (e.g.\cite{Mar95cc}),
and harmonic frequencies (e.g. \cite{Mar95cc}). The treatment of
inner-shell correlation requires special basis sets which not only
have additional radial flexibility, but also include `hard' or `tight'
(i.e., high-exponent) $d$,$f$,\ldots{} functions in order
to cover angular correlation from the inner-shell orbitals. Special
basis sets of this type have been developed, such as
the cc-pCV$n$Z (correlation consistent polarized core-valence
$n$-tuple zeta) basis sets of Woon and Dunning\cite{cc-pCVnZ}
for B--Ne, the MT (Martin-Taylor\cite{MTbasis}) and MTsmall\cite{W1}
basis sets for Li--Ar, the rather small G3large basis set\cite{G3theory}
of Pople and coworkers for the main group elements of the first three
rows, and very recently the cc-pwCV$n$Z (i.e. core-valence weighted
cc-pCV$n$Z) basis sets of Peterson and Dunning\cite{cc-pwCVnZ} 
for B--Ne and Al--Ar.

While the CC+CV contribution to absolute correlation energies may
rival or exceed the valence contribution, its differential contribution
to the molecular binding energy is generally small compared to the
SCF and valence correlation contributions (typically less than 1\%
for 1st-row systems). For that reason --- as well as the formidable
cost of core correlation calculations --- computational thermochemistry
schemes (e.g. G3 theory\cite{G3theory}, W1/W2 theory\cite{W1,W1chapter})
that account for inner-shell correlation generally treat the latter
as a small additive contribution obtained with a relatively compact
core correlation basis set. (The finding\cite{W1} that connected
triple substitutions are surprisingly important for CV contributions
ensures that the core correlation step often is the rate-determining
one for benchmark thermochemistry calculations, particularly beyond
the first row of the Periodic Table.) Recently, some attempts have
been made to replace the core correlation step by bond additivity
approximations\cite[and references therein]{MSFT}, and promising
results have been obtained\cite{CPPon1stRow} by means of core polarization
potentials\cite{CPP}.

Despite repeated warnings in the literature against the practice (e.g.,
by Taylor\cite{Taylor92lnqc}, by one of the present authors\cite{Mar95cc},
and by Woon and Dunning\cite{cc-pCVnZ}), core correlation is often
included --- for technical reasons or `because it cannot hurt' ---
in calculations using basis sets (e.g. the standard Dunning correlation
consistent basis sets\cite{Dun89,DunCCreview}, which are of minimal basis set
quality in the inner-shell orbitals) which are not adapted for nonvalence
correlation. This can often cause errors on computed properties well
in excess of the basis set incompleteness error: For instance, a
comparison between valence-only\cite{Mar96c4} and all-electron\cite{Sta92c4}
CCSD(T)/cc-pVTZ harmonic frequency calculations on the cyclic isomer
of C$_{4}$ suggests core correlation contributions to the harmonic
frequencies of up to 50 cm$^{-1}$, a factor of five more than
the true correction obtained\cite{Mar96c4} with a core correlation
basis set.

For the elements B--Ne and Al--Ar, the gap between valence and inner-shell
orbital energies is large enough that a conventional core-valence
separation of the orbital energies is easily made. Things are less
simple further down the periodic table. Bauschlicher\cite{BauGa,BauIn}
noted that for gallium and indium halides, the valence (2s) orbitals
of the halogen are below the Ga(3d) or In(4d) `core' orbitals in energy.
At the very least, these d orbitals should be added to the correlation
space; suitable basis sets for this purpose have been developed by
Martin and Sundermann\cite{SDB-cc} for Ga and Ge, and by Bauschlicher
\cite{BauIn} for In.

But similar issues arise with Group 1 and 2 elements. Radom and
coworkers\cite{Radom99,Radom03} recently noted catastrophic failures of G2
theory and related methods in predicting the heats of formation of
various alkali and alkaline earth metal oxides and hydroxides: for example,
the binding energy of K$_{2}$O
is underestimated by no less than 256 kJ/mol! Upon correlating all
electrons, this error decreases to 60 kJ/mol, which is further decreased
to 17 kJ/mol when more sophisticated electron correlation methods
are used. Inspection of an atomic orbital energy table\cite{atenertab}
quickly reveals why: the oxygen 2s and 2p valence orbitals lie below
the potassium 3s and 3p inner-shell orbitals, rendering the conventional
separation between `valence' and `core' orbitals essentially meaningless.
Inclusion of Group 1 and 2 subvalence correlation --- termed `relaxed
inner valence' (RIV) by Radom and coworkers\cite{Radom99} --- may
in fact be the appropriate treatment, not just for K (and Ca) compounds,
but for alkali and alkaline earth metal compounds in general. (One
of the first papers to recognize the importance of inner-shell correlation
for alkali metals may have been that by Liu and coworkers.\cite{Liu83})

This problem is far from academic, considering the great importance
of K$^{+}$, Ca$^{2+}$, Na$^{+}$, and Mg$^{2+}$
complexes in molecular biology. In addition, a number of these interactions
(e.g., of the cation-$\pi$ type) are not necessarily amenable
to density functional treatments without validation of some kind by
means of ab initio methods. Clearly, the availability of high-quality
core-valence basis sets for the elements Li, Be, Na, Mg, K, and Ca
would benefit the high-accuracy computational thermochemistry and
spectroscopy communities as well as biomolecular modelers. The purpose
of the present work is the development and validation of CVnZ (core-valence
n-tuple zeta, n=D, T, Q, 5) basis sets for these elements.

\section{Computational details}

All electronic structure calculations carried out in the present work
were carried out using the MOLPRO 2000.1 and 2002.3 program systems\cite{M2K,M2002}
running on SGI Origin 2000 (12 $\times $MIPS R10000, 300 MHz
and 4 $\times $MIPS R10000, 195 MHz, IRIX 6.5) and Compaq ES40
(4 $\times $ EV67, 667 MHz, Tru64 Unix 4.0f) minisupercomputers,
as well as Compaq XP1000 (EV6, 500 MHz, Tru64 Unix 4.0f) and dual Intel
Xeon (1.7, 2.0 and 2.4 GHz, RedHat Linux 7.2 and 7.3) workstations.

Unless explicitly noted otherwise, energy calculations were carried
out at the CCSD(T) level; for open-shell systems, single-determinant
ROHF reference functions were used, but the definition of the CCSD(T)
energy according to Watts et al.\cite{Wat93} was employed throughout.

Validation calculations on diatomic molecules were carried out as
follows. Energies were computed at 21 equidistant points around the
putative bond distance with interpoint spacings of 0.01--0.03~\AA,
depending on the curvature of the surface. Energies at these points
were converged as precisely as feasible, with integral evaluation
thresholds being tightened as necessary. Polynomials of increasing
order were then fit through the points, and the significance of the
additional order terms subjected to a Fisher-Snedecor test. The expansion
was truncated at the point where the two-way significance of the next
higher order dropped below 99\%. The retained expansion was generally
of order between six and eight. The minimum of this curve was then
sought by means of a Newton-Raphson method, and the curve re-expanded
around the minimum. A Dunham analysis\cite{Dun32} was then carried out on the
final curve. The computed dissociation energies at 0~K, $D_0$, reported in the
tables include anharmonic zero-point corrections from the computed 
$\omega_e$ and $\omega_ex_e$ at the same level of theory.

As for the accompanying basis sets, standard cc-pV$n$Z basis
sets were used throughout on H. For O and F, aug-cc-pV$n$Z basis
sets\cite{AVnZ} were used in valence calculations and aug-cc-pCV$n$Z
basis sets\cite{cc-pCVnZ} in calculation where the O and F (1s) cores
were correlated. For S and Cl, the aug-cc-pV$(n+d)$Z basis sets
of Wilson et al.\cite{Wilson2001} were used in valence calculations,
and the cc-pCV$n$Z basis sets of Peterson\cite{cc-pwCVnZ} in
calculations where the (2s,2p) cores of these elements were correlated.

Where deemed necessary, scalar relativistic effects were assessed by
means of CCSD(T) calculations within the Douglas-Kroll-Hess approximation\cite{DK,Hes86}
as implemented in MOLPRO 2002.3\cite{M2002}.

\section{Optimization of CV\protect\protect$\lowercase {n}\protect \protect$Z
basis sets}

\subsection{General procedure}

The basis set optimizations were carried out using a Fortran
program developed in-house. Derivatives were obtained numerically
using central differences, yielding both the gradient and the diagonal
of the Hessian. The step thus obtained --- essentially Newton-Raphson
with neglect of off-diagonal Hessian elements --- was combined with
a Brent line search. Our experience with this simple but fairly robust
optimization algorithm is quite good as long as optimization parameters
are not too strongly coupled. In the presence of significant coupling,
DIIS (Direct Inversion of the Iterative Subspace) \cite{diis} dramatically
sped up convergence. 

Finite difference step sizes were made
roughly proportional to the parameters themselves, and increased as
necessary if the parameter surface in the affected direction was found to be
particularly `flat'.

Some additional optimizations were carried out using an adaptation
of the DOMIN program by P. Spellucci\cite{domin}, which is an implementation
of the BFGS (Broyden-Fletcher-Goldfarb-Shanno) variable-metric method.
Numerical derivatives of order two, four, and six were used: the lower
orders until an approximate minimum was reached, after which the optimization
was refined using the higher orders. In most cases, however, the simple
procedure outlined above appeared to be more robust.

For the purposes of the optimization (and since only very small systems
are involved), integral evaluation, SCF and CI convergence thresholds
in MOLPRO were tightened to essentially machine precision. Basis set
parameters were converged to at least four significant digits, and
five where at all possible.

In cases where many primitives of a particular angular symmetry are
required, we initially constrained these functions to follow an even-tempered
sequence $\zeta_{i}=\alpha \beta ^{(i-n-1)/2}$, where $n$
is the total number of primitives. Only when optimum $\alpha $
and $\beta $ had been reached were the individual exponents optimized
further without any constraints. If a set of $(n-1)$ primitives
of that symmetry was already available (\emph{in casu}, from a previously
optimized smaller basis set), the geometric mean of their exponents
was taken as the starting value for $\alpha $, and the arithmetic
mean of the ratios between successive primitives as the initial value
for $\beta $.

A few `checks and balances' can generally be applied to the final exponents: 

\begin{enumerate}
\item within a row, the geometric mean of the exponents should increase
roughly as $(Z^{*})^{2}$, where $Z^{*}$ is the `shielded'
nuclear charge (first Ahlrichs-Taylor rule\cite{Ahl80}); 
\item exponents in a larger set (e.g. $3d$) should mesh with the next
smaller set (e.g. $2d$);
\item the ratio of successive exponents in a set should be about two or higher,
and all gaps within the core or valence parts should roughly be of
the same order;
\item the geometric mean of the exponents for each angular momentum should
roughly increase by a factor of 1.2 for each step up
in $L$ (second Ahlrichs-Taylor rule\cite{Ahl80}).
\end{enumerate}

\subsection{Li and Be}

For Li and Be, the `unofficial' cc-pV$n$Z basis sets of Feller\cite{Feller}
supplied in the MOLPRO 2000.1 basis set library were taken as starting
points. The quantity optimized for is the (1s) correlation energy
at the CISD level for the Li$(^{2}S)$ and Be$(^{1}S)$ ground
states, respectively. This happens to be equivalent to $E${[}CISD,full{]}$-E${[}SCF{]}
for Li, but is equal to $E${[}CISD,full{]}-$E${[}CISD,valence{]}
for Be.

There exists a tradition of optimizing basis sets for correlated calculations
at the CISD level, since CISD is a variational method. However, optimizations
for the CCSD or CCSD(T) counterparts of the abovementioned correlation
energies yield very similar basis sets, which would have been of identical
quality in molecular calculations. (A table of these exponents for the 
CVDZ, CVTZ, and CVQZ cases can be found in the Supplementary Material\cite{suppmat}.)

By analogy with the cc-pCV$n$Z basis sets for B--Ne\cite{Dun95},
$1s1p$, $2s2p1d$, $3s3p2d1f$, and $4s4p3d2f1g$
sets of primitives were added to the cc-pVDZ, cc-pVTZ, cc-pVQZ, and
cc-pV5Z basis sets, respectively. Successive angular momenta were
optimized individually, after which the complete set of exponents
was further optimized together.

Multiple local minima exist for the larger sets, and care was taken
to ensure that the final optimized basis set reflects the most `contiguous'
(or most `even-tempered') solution in the sense of having no obvious
`gaps' between exponents.

Aside from this issue, the basis set optimizations proceeded uneventfully.

\subsection{Na and Mg}

Likewise, for Na and Mg, unpublished cc-pV$n$Z basis sets of
Feller\cite{Feller} were taken as the starting point. In this case,
we optimized for the $2s2p$ correlation energy, found as
$E${[}CISD,2s2p3s{]}-$E${[}CISD,3s{]} for Mg and as simply
$E${[}CISD,2s2p3s{]}-$E${[}SCF{]} for Na. Note that the
$1s$ orbitals are constrained to be doubly occupied throughout:
not only will the effect of this 'deep core' orbital on chemically
significant properties be negligible, but including it in the optimization
would bias the exponents towards the large --- but chemically quite
invariant --- core-core correlation energy rather than the chemically
more significant $2s2p$ core-core and core-valence correlation
energies.

Since the highest core correlation orbital is of $p$ symmetry
in this case, the basis sets are rather larger, involving addition
of $1s1p1d$, $2s2p2d1f$, $3s3p3d2f1g$, $4s4p4d3f2g1h$
core correlation sets to the cc-pVDZ, cc-pVTZ, cc-pVQZ, and cc-pV5Z
sets, respectively.

For the CV5Z basis sets for Na and Mg, simply optimizing a set of core correlation
$s$ functions leads to such serious near-linear dependence problems
that the basis would in practice be unusable. For want of an alternative,
we simply uncontracted an additional four $s$ primitives instead.

\subsection{K and Ca}

The core-valence gap in Ca, and particularly K, is so small that the
optimization of regular cc-pV$n$Z basis sets would be of largely
academic interest. Feller and coworkers previously published CVDZ,
CVTZ, and CVQZ basis sets for K\cite{K-CVDZ}; the $sp$ exponents
for the CVDZ and CVTZ basis sets were taken from (15s,12p) and (18s,15p)
basis sets, respectively, of Ahlrichs and coworkers\cite{Ahl92},
while those for the CVQZ basis set were taken from the Partridge-1
set\cite{Partridge}, which is of (23s,19p) uncontracted size.

Initially we employed these basis sets unaltered, and merely added 
a CV5Z basis set. This was obtained as follows, starting from the
Partridge-3 basis set:
\begin{itemize}
\item the basis set was contracted to miminal (i.e. $[4s3p]$) in an atomic
calculation on the ground state
\item the four outermost $s$ and $p$ primitives each were decontracted
\item $[4d3f2g1h]$ functions were optimized at the valence-only CISD level
for the K$_2$ molecule at its experimental bond distance, yielding an intermediate
cc-pV5Z basis set
\item core correlation functions were obtained by optimizing the 
atomic $E$[CISD,3s3p4s]-$E$[SCF] energy. The optimization of the $s$ and $p$ 
functions was plagued by insurmountable near-linear dependence problems, and we ended up merely
decontracting an additional four primitives of $s$ and $p$
symmetries each. In addition, a crossover occurred between the highest
$d$ exponent of the underlying cc-pV5Z basis set and the lowest
$d$ exponent required for the inner-shell correlation, and it
was decided to merely keep the outermost three $d$ primitives
constant while optimizing all five remaining $d$ primitives for
$3s3p$ correlation. 
\end{itemize}

Thus, our original CV5Z basis set was obtained.
However, as evidenced in Table \ref{tab:row3}, 
basis set convergence of properties 
for K$_2$ and KH was less than satisfactory. In particular,
inspection of the dissociation energy as a function of the basis set
reveals that for K$_2$, the CVTZ, CVQZ, and CV5Z results are nearly
collinear, which is obviously undesirable. For this reason, the $d,f,\ldots$
exponents of the K CVTZ and CVQZ basis sets were reoptimized in the same manner
as the CV5Z basis set. In the process we found the `valence correlation' $d$ and 
$f$ exponents in the original Feller basis sets to be excessively `tight', and
with the adjusted basis sets (denoted `version 0.1' in Table \ref{tab:row3}), 
increases of 6.5 and 7.0 kJ/mol are seen in the
CVTZ and CVQZ $D_0$ values. Other molecular constants are also affected quite
significantly. While the Feller CVTZ and CVQZ were not found to be as problematic
for the other K compounds considered in this work, they are clearly unsuitable
for K$_2$ or any other system with similar long-distance bonding.

Analogously, we optimized CV$n$Z ($n$=D,T,Q,5) basis sets for Ca, starting 
from Ahlrichs $14s9p$, Partridge-1 $20s12p$, Partridge-2 $23s16p$, and
Partridge-3 $26s16p$ primitive sets. Initial attempts to optimize basis sets for the
$^3P$ atomic excited state (analogously to the sp-hybridized state of Be)
yielded exceedingly poor basis set convergence of molecular properties, and 
these basis sets were abandoned in favor of $^1S$ ground-state optimized basis sets. 
(We also attempted optimizations in the CaH$_2$ molecule, and found basis set
parameters obtained there to be very similar to those for the $^1S$ atomic
ground state.) Valence correlation functions were optimized for the valence
correlation energy of the ground-state calcium atom, and core-valence functions
for $E${[}CISD,3s3p4s{]}-$E${[}CISD,4s{]}. For the CVQZ and CV5Z 
basis sets, the optimized `valence' and `subvalence' $d$
and $f$ shells interlock, and as a result only the outermost
$2d2f$ were held at their valence-optimized values and the remainder
reoptimized for inner-shell correlation. No such issues arose with
the $g$ and $h$ functions.
Once again, basis set convergence of molecular properties (e.g. for CaH, but
also for other diatomics, not displayed here)
was found to be somewhat 
unsatisfactory.

For both atoms, we then optimized a sequence of four `stretch-tuned' 
basis sets\cite{stretch}, i.e. basis sets in which the exponents of each angular 
momentum obey the following four-parameter relation:
\begin{equation}
\ln \zeta_{n}=\alpha +n(\beta +(n-1)(\gamma +(n-2)\delta ))
\end{equation}
(A total of only eight parameters thus had to be optimized for each basis set.)
The particular sequence of contraction sizes chosen was 
$(20s12p)$, $(22s14p)$, $(24s16p)$, and $(26s18p)$. The valence and inner-shell
correlation functions were optimized as before.

These basis sets (denoted `version 0.2' in Table \ref{tab:row3}) appeared to have better convergence properties.
However, something was clearly still not satisfactory. For instance, basis
set convergence for the ionization potentials of K and Ca atoms was found to 
be atypically slow (Table \ref{tab:IPs}). 

It then occurred to us that, as the $4s$ orbital of these atoms is very diffuse, the 
atom-optimized $p$ functions may be too `tight' (as they were optimized for atoms with
vacant $4p$ orbitals) to contain suitable first polarization/angular correlation functions
for the outer part of the $4s$ orbital. This assumption was verified by optimizing single `probe' basis 
functions of various symmetries on top of CV$n$Z basis sets held constant. (Only valence electrons
were correlated, as otherwise the functions would have fallen into the `gravity well' of the
inner-shell electrons.) A very noticeable lowering of the total energy was seen upon adding a 
single $p$ function, the optimum exponent of which indeed turned out to be `looser' than any
of the existing $p$ functions. An additional `loose' $p$ function led to a further improvement for the K
CV$n$Z ($n$=D,T,Q) basis sets, but not for the K CV5Z basis set (the outer $p$ primitive of which is
already quite `loose') or any of the Ca basis sets.
Consequently, the valence correlation
parts of the CV$n$Z basis sets were reoptimized with two and one
additional $p$ functions added, respectively, for K and Ca, and afterwards the 
inner-shell part was reoptimized for consistency. Thus our final CV$n$Z basis 
sets were obtained. As seen in Table \ref{tab:row3}, basis set convergence of molecular parameters
is now satisfactory. For IP(K) and IP(Ca), nearly exact values are now obtained (Table \ref{tab:IPs}). 

Since the primitive K and Ca basis sets may be somewhat bulky for some applications,
we have also generated SDB-CV$n$Z basis sets, which are combinations of the
valence and subvalence parts of the CV$n$Z basis sets with small-core
Stuttgart-Dresden-Bonn\cite{SDBreview} relativistic effective core
potentials. The latter replace the (1s2s2p) electrons. Unlike the situation
with large-core basis set where reoptimization of the basis set is essential\cite{SDB-cc},
we have kept all exponents at their all-electron values and merely
`pruned' and recontracted the basis set. Specifically, the atomic SCF calculation
was repeated with the (1s2s2p) orbitals replaced by the SDB pseudopotential
denoted `ECP10MWB'. Then all primitives with absolute coefficients of less than
10$^{-5}$ in any valence orbitals were deleted, and the orbitals recontracted.

All basis sets obtained in this work are available in machine-readable form
(Gaussian and MOLPRO formats) as supplementary material\cite{suppmat}
to the present paper. Contracted basis set sizes and numbers of basis functions
for each element are given in Table \ref{tab:contract}.

\section{Validation for diatomic molecules}

\subsection{Diatomic metal hydrides}

Since hydrogen has no core electrons, there are no differential core-core
contributions (just core-valence) to the molecular properties. Computed
and experimental data for the diatomic hydrides are given in Table \ref{tab:hydrides}.

As noted previously, the inner-shell contribution to the LiH spectroscopic
constants is quite appreciable, as can be expected from the small
Li(1s)-H(2s) gap of about 2 a.u. We note that the CCSD(T,riv)/CV5Z
results reproduce the Born-Oppenheimer bond distance and harmonic
frequency to within 0.0002~\AA\ and 0.2 cm$^{-1}$, respectively.
Neglect of Li(1s) correlation leads to errors of 0.012~\AA\ and 13.5
cm$^{-1}$, respectively. Nevertheless, the contribution to D$_{\textrm{e}}$
does not exceed 1.2 kJ/mol. 

The corresponding core-valence gap for BeH is almost doubled, which
translates into substantially reduced core-valence contributions (relatively
speaking) to $r_{e}$ and $\omega_{e}$. Interestingly,
the contribution to $D_0$ reaches nearly 2 kJ/mol.

In contrast, the Na(2p)-H(1s) gap narrows to no more than 1 a.u.,
and in NaH we see inner-shell correlation contributions of 33 cm$^{-1}$
to $\omega_{e}$ and 0.035~\AA\ to $r_{e}$. Interestingly,
once more $D_{e}$ is nearly unaffected. Once more, the CCSD(T,riv)/CV5Z
results are in excellent agreement with experiment for the spectroscopic
constants: applying a W2-type extrapolation for $D_0$ results in a value
of 182.35 kJ/mol, within 0.3 kJ/mol of experiment. 

Given that the Mg(2p)-H(1s) gap is not dissimilar from the corresponding
one in LiH, it is not greatly surprising that the importance of core-valence
contributions is again much smaller than in NaH. However, the 8 kJ/mol
contribution to $D_0$ is considerably more significant.

In KH, the K(3p) and H(1s) orbitals are closer than 0.5 a.u., and
the contributions of nearly 0.1~\AA\ and 44 cm$^{-1}$ to $r_{e}$
and $\omega_{e}$, respectively, clearly suggest that any sort
of `valence correlation only' calculation on a K compound should be
viewed with great suspicion. Note that the core-valence contribution
to $D_0$ is \emph{negative} in this case, as was previously found (e.g.
\cite{MSFT}) for aluminum and silicon hydrides. 

In comparison, the Ca(3p)--H(1s) gap widens to 0.84 a.u., and core-valence
effects on the molecular properties are mitigated accordingly. Like for 
MgH and KH, inner-shell correlation in fact slightly reduces the
dissociation energy.

With the exception of KH and CaH, the differential core-valence
contributions appear to have converged with respect to the basis set
at the CVQZ stage. This in itself satisfies a minimum requirement
for treating the inner-shell correlation contribution separately,
something which is definitely inappropriate for KH.

On the whole, agreement between CCSD(T,riv)/CV5Z and experiment 
can only be described as quite satisfactory. For systems other than 
LiH and NaH, imperfections in the CCSD(T) electron correlation method
may account for most of the remaining discrepancy between computed and
observed harmonic frequencies\cite{ch}.

\subsection{Metal diatomics}

The alkaline earth metal dimers (particularly Be$_2$) exhibit
such severe multireference character that they warrant studies in
themselves (which would, however, focus on electron correlation methods
rather than basis sets, see e.g.\cite{Be2} and references therein). 
Results for species other than Be$_2$ can be found
in Table \ref{tab:dimetals}.

In Li$_{2}$, subvalence correlation accounts for 0.024~\AA, which
is sizable by spectroscopist's standards. In Na$_{2}$, this is
drastically increased to nearly 0.1~\AA, and reaches a whopping 0.23~\AA\ 
in K$_{2}$. Interestingly, these contributions are quite close
to double their metal hydride counterparts. As expected, LiNa and
NaK represent scenarios intermediate between the homonuclear diatomics
of the constituent elements. Changes in $\omega_{e}$ are modest
in absolute numbers, but for these molecules represent relative errors
of up to 5\%. It is quite clear however that, with the possible exception
of Li$_{2}$, inclusion of subvalence correlation is essential
for reliable molecular parameters. Once again we see, however, that
bond energies are only affected mildly: 0.8 kJ/mol in Li$_{2}$,
1.7 kJ/mol in LiNa, 1.6 kJ/mol in Na$_{2}$, and only 0.5 kJ/mol
in K$_{2}$. 

Once again, a basis set of CVQZ quality appears to be close to convergence
for the differential core-valence effects. 

Effects for Mg$_2$ and Ca$_2$ are considerably milder than for their
alkali neighbors, yet $(3s,3p)$ correlation reduces the Ca$_2$ bond
distance by 0.07--0.08~\AA, not negligible by any reasonable standard.
The discrepancies between theory and experiment for Mg$_2$ primarily reflect 
the inadequacy of the CCSD(T) method for this highly multireference system.
In contrast, Ca$_2$ yields quite satisfying results compared to experiment.

Agreement between our best calculations and experiment can only be described
as excellent for Li$_2$. The same would be true of Na$_2$ if it were not for
the bond length which is still 0.003~\AA\ too long at the CCSD(T,riv)/CV5Z
level. However, at the DK-CCSD(T)/CVQZ level,
we find a scalar relativistic
correction to the bond length of -0.006~\AA.
(If this correction would seem to be exaggerated, 
we note that the Na$_2$ potential curve is quite flat, and that 
already for OH$^-$ and HF, scalar relativistic corrections were found to be required
for spectroscopic-quality results\cite{OH-}.) In K$_2$, our best calculated
results are likewise in excellent agreement with the experimental spectroscopic
data except for $r_e$, which is calculated to be 0.014~\AA\ longer than
the observed value. 
At the DK-CCSD(T)/CVQZ level,
we find a relativistic
contribution of -0.013~\AA\ to the bond length, which explains most of the
discrepancy. In addition, the $(2s,2p)$ `deep core' orbitals are energetically
in the same range as the $(1s)$ orbitals in C and N, so it cannot be entirely ruled out
that `deep core' correlation may affect molecular properties in K
compounds.

We optimized a CCVTZ (deep core valence triple zeta) basis set
for K. 
In these calculations, the CVTZ basis set was held constant and
exponents of $2s2p2d1f$ basis functions 
optimized for $E$[CISD,2s2p3s3p4s]$-E$[CISD,3s3p4s], i.e. for deep-core
correlation energy taken in isolation. As can be seen in Table \ref{tab:exponents}, 
these exponents are obviously much `tighter', by almost an order of magnitude,
than those required for subvalence correlation. Hence, any `deep core'
result in a mere valence and subvalence CV$n$Z (let alone a valence-only
cc-pV$n$Z set) should be regarded with skepticism at best.

 The only
change of note we see in the molecular properties when using this CCVTZ basis set
for K$_2$ is a shortening of the
K--K bond by another 
0.0018 \AA.
We also attempted to optimize a CCVQZ basis set
but ran into insurmountable near-linear-dependence problems; however, the
CCVTZ result should at least give an indication of the (fairly modest) 
magnitude of this effect.

\subsection{Diatomic metal halides}

This set of molecules is particularly revelant because of the deep-lying
valence $s$ orbitals on the halogen atoms. Relevant results are
collected in Table \ref{tab:halides}.

Firstly, for LiF the contribution of Li(1s) correlation to the binding
energy is somewhat more noticeable (4 kJ/mol). Its inclusion also
causes a contraction of the bond by 0.015~\AA\ and an increase in $\omega_{e}$
of 12 cm$^{-1}$. The further effect of permitting F(1s) correlation
is nearly negligible in comparison. The Cl(2s,2p) correlation effect
is somewhat more noticeable but still Li(1s) correlation accounts
for the lion's share of the changes. 

The F(2s) and Na(2p) orbitals are nearly degenerate, and hence one
would expect valence-only calculations on NaF to be problematic, to
say the least. Nevertheless, while the effect of Na(2s,2p) correlation
on the molecular constants is quite noticeable and its inclusion clearly
essential for accurate calculations, the results obtained are not
outright nonsensical. Of course, considering the difference in electronegativity
between the elements, the combination of Na$^{+}$ and F$^{-}$
would yield a more appropriate (if perhaps extreme) `atoms in molecules'
picture --- and the corresponding orbital energy gap in this admittedly
extreme scenario is 0.72 a.u. with the Partridge 3 basis set. The
corresponding `ionic' gap for NaCl amounts to 1.07 a.u., explaining
why also for NaCl, valence-only results are not totally unreasonable. 

The person carrying out calculations on KF and KCl has no such luck. The
inclusion of K(3s3p) correlation shifts $\omega_{e}$ in KF upwards
by no less than 20\%, and the contribution to $D_{e}$ is seen
to be on the order of 170 kJ/mol. In fact, the `valence' results are
better described as referring to correlating the \emph{z} component
of K(3p) and freezing the F(2s), which obviously makes no chemical
sense at all. 
The valence-computed dissociation energy of KCl agrees deceptively
well with experiment, but an error of 0.3~\AA\ in the bond distance 
and of a factor of three in the anharmonicity constant should discourage
any quantum chemist from performing valence-only calculations on this type of
species.

While obviously significant for spectroscopic purposes, the effects
of metal subvalence correlation in MgF and MgCl are clearly less prominent
than in NaF and NaCl, respectively, and with the beryllium halides, one
clearly could `get away' with a differential treatment. In contrast,
in CaF a calcium subvalence contribution to $D_0$ of 180 kJ/mol 
is found, not to mention +46 cm$^{-1}$ on $\omega_e$ and a factor
of two on $\omega_ex_e$. Oddly, the effect on the bond length is 
comparatively small (0.04~\AA). The opposite scenario is seen for
CaCl, with a $\Delta r_e$ of 0.1~\AA\ but otherwise not outlandishly
erroneous values for the other spectroscopic constants.
CaF and CaCl results resemble more the potassium halides. 

And once again, the CCSD(T,riv)/CV5Z results agree as well with experiment 
as can reasonably be expected.

We have also considered calculations in which subvalence correlation
on the halide was permitted. However, for K and Ca compounds with
\{C, N, O, F\}, the (1s) orbital of these latter elements is in fact below the 
K(2s,2p) `deep core' orbitals in energy, and therefore any such calculation
will require the use of CCV$n$Z `deep core' basis sets on the metal.
Such calculations are sufficiently costly that one might want to
limit them to single-point energy calculations in high-accuracy
thermodynamic work. For MCl, the most noticeable effect of including
$(2s,2p)$ subvalence correlation on chlorine is a contraction of $r_e$
by 0.003~\AA.

\subsection{The metal chalcogenides}

We will primarily focus on the oxides. Results can be found in Table \ref{tab:oxides}.

Clearly, changes of 13 cm$^{-1}$ and 6 kJ/mol, respectively, for $\omega_{e}$
and $D_{e}$ of BeO suggest the importance of metal (1s) correlation
in this system; contributions from O(1s) correlation are an order
of magnitude less important. Note that the experimental data can be
reproduced almost exactly at the CCSD(T,all)/CV5Z level, despite BeO
having pronounced multireference character and in fact being on the
borderline of applicability of CCSD(T) (see e.g.\cite{Urban}). 
In the isovalent BeS system, Be(1s)
correlation is still noticeably more important than S(2s,2p) correlation,
although the difference is not as pronounced as its counterpart in
BeO.

Changes between valence-only and RIV treatment 
of 0.017~\AA\ in $r_e$(MgO) and 23 cm$^{-1}$ in $\omega_{e}$(MgO)
speak for themselves. The MgO molecule has
very pronounced multireference character, and accurate reproduction
of the experimental spectroscopic constants 
would required an elaborate multireference treatment.
Subvalence correlation effects in MgS are rather
milder.

In contrast, inspection of the computed spectroscopic constants for
CaO immediately reveals that valence-only calculations on this system
are essentially exercises in wasting computer time. Valence-only bond
distances and dissociation energies are off by a quarter of an \AA, and
a factor of six in $D_0$.
Respectable agreement
with experiment can however be reached by correlating the Ca(3s,3p)
electrons. Obviously, the large uncertainties in the measured dissociation 
energies preclude really fine comparisons with experiment: on the basis of the
preceding, it can perhaps be stated that the calculated binding energies are
more reliable than the experimental ones for the alkali(ne earth) metal
chalcogenides.

Once again, correlating the $(1s)$ orbital on O would require including 
$(2s,2p)$ deep-core correlation in Ca, otherwise meaningless results are obtained.

\subsection{Additional $d$ functions on Ca}

Wesolowski et al.\cite{Fritz}, in a study of calcium oxide, observed especially poor
performance for a valence-only atomic natural orbital (ANO) basis set\cite{Roos}. This was
considerably improved when high-exponent $d$ functions were added. They noted that
Ca$^+$ has a low-lying $^2D$ state only 1.7 eV above the $^2S$ ground state, and showed
that said ANO basis set overestimates this separation by nearly a factor of three.
Addition of several very high-exponent $d$ functions, taken from the 6-311G(2df) basis
set for Ca\cite{Blaudeau} (which treats the $(3d)$ orbital on a valence footing), 
results in an at least semiquantitatively correct result.

As the present basis sets were developed from the ground up with inner-shell correlation
in mind, they intrinsically contain high-exponent $d$ functions and should therefore 
be much less susceptible to the problem. As shown in Table \ref{tab:Ca+}, quite good agreement
with the experimental transition energy is obtained for CVQZ and CV5Z basis sets, but 
the smaller basis sets leave something to be desired. 

We therefore proceeded to optimize CV$(n+d)$Z, CV$(n+2d)$Z, and CV$(n+3d)$Z basis sets
in the following fashion. (Both the nomenclature and the procedure bear resemblance to the
cc-pV$(n+d)$Z basis sets for second-row elements\cite{Wilson2001}.) All angular momenta
other than $d$ were kept constant, as were the valence-optimized $d$ functions (see above).
The remaining $d$ functions were reoptimized for inner-shell correlation with one, two, or
three additional $d$ functions added. (In the CV(Q+d)Z and CV(Q+2d)Z basis sets, it was 
found necessary to constrain the exponents to stretch-tuned and even-tempered sequences,
respectively.)

As can be seen in Table \ref{tab:Ca+}, the $^2D\leftarrow{}^2S$ excitation energy indeed
converges considerably faster for the CV$(n+d)$Z and CV$(n+2d)$Z series than for the CV$n$Z
series. After applying a scalar relativistic correction by means of the Douglas-Kroll method
in an uncontracted CV(Q+d)Z basis sets, very good agreement with experiment can be achieved.
As for the molecular properties of CaO (Table \ref{tab:oxides}), the effect of additional
$d$ functions is quite dramatic in the CVDZ case and still quite important in the CVTZ
case, but tapers off for CVQZ and becomes quite insignificant for CV5Z. We noticed similar
behavior in the other CaX systems considered in this paper (Tables \ref{tab:hydrides},
\ref{tab:dimetals}, and \ref{tab:halides}). Convergence as a function of the number of
additional $d$ functions appears to be approached for CV(D+3d)Z, CV(T+2d)Z, CV(Q+d)Z,
and CV5Z (note particularly Table \ref{tab:oxides}). Basis set convergence of molecular
properties along this sequence is clearly much smoother than for the unmodified CV$n$Z
basis sets. This is particularly true for CaO and, to a somewhat lesser extent, for
CaF, least so for CaH and Ca$_2$ where basis set convergence was adequate to begin with.

\section{Conclusions}

Core-valence basis sets have been developed for the alkali and 
alkaline earth metals Li, Be, Na, Mg, K, and Ca. Validation 
calculations for a number of diatomics involving these systems reveal 
basis set convergence to be satisfactory.

The addition of low-exponent $p$ functions to the K and Ca basis set
is found to be essential for smooth basis set convergence
of molecular properties. These functions accommodate angular correlation
from the outer part of the $4s$ orbital.

Valence-only calculations on K and Ca chalcogenides and halides yield
large errors at best, and chemically nonsensical results at worst.
In general, inclusion of subvalence 
correlation in K and Ca compounds is absolutely essential for even
\emph{reliable} (let alone \emph{accurate}) results, while for accurate
calculations, subvalence correlation in Na (and, to a lesser extent,
Li and Mg) should be included, preferably as a `baseline' treatment
rather than an additive correction. The latter appears to be sufficient
for Be compounds. Any study that seeks to address the importance of
first-row element core correlation in K and Ca compounds should
also consider deep-core (2s,2p) correlation from these latter elements.

For compounds involving Ca bound to highly electronegative elements, we
strongly recommend basis sets with additional $d$ functions on Ca: our
suggested sequence for basis set convergence studies would be 
CV(D+3d)Z, CV(T+2d)Z, CV(Q+d)Z, CV5Z.

\begin{acknowledgments}
JM is a member of the Lise Meitner Center for Computational Quantum Chemistry (Israel-Germany)
and the Helen and Martin Kimmel Center for Molecular Design (Weizmann Institute).
This research was supported by the \textit{Tashtiyot} program of the
Ministry of Science and Technology (Israel) and by the Minerva Foundation, Munich, Germany.
MAI and MO acknowledge Ph.D. and M.Sc. Fellowships, respectively, from the
Feinberg Graduate School (Weizmann Institute of Science). We thank a referee for bringing
Ref.\cite{Fritz} to our attention.

\end{acknowledgments}

\clearpage
\begingroup
\squeezetable
\begin{table}[h]
\caption{Comparison of different CV$n$Z basis sets for properties of K$_2$ and KH\label{tab:row3}.}

\begin{tabular}{lrrrr}
\hline\hline
& $r_e$ & $\omega_e$ & $\omega_ex_e$ & $D_0$ \\
& (\AA) & (cm$^{-1}$) & (cm$^{-1}$) & (kJ/mol) \\
\hline
\multicolumn{5}{c}{KH}\\
\hline
CVDZ Feller     & 2.2835 &  935.7  & 13.80 & 124.3 \\
CVDZ version 0.1& 2.2847 &  930.0  & 13.87 & 123.2 \\
CVDZ version 0.2& 2.2673 &  960.5  & 18.61 & 125.4 \\
Final CVDZ (incl. +2p)& 2.3232 &  902.1  & 11.42 & 136.1 \\
CVTZ Feller     & 2.2382 &  998.8  & 16.69 & 156.7 \\
CVTZ version 0.1& 2.2410 &  991.0  & 17.11 & 153.0 \\
CVTZ version 0.2& 2.2360 & 1025.3  & 13.64 & 159.5 \\
Final CVTZ (incl. +2p)& 2.2591 &  976.3  & 15.41 & 161.2 \\
CVQZ Feller     & 2.2353 & 1002.9  & 14.24 & 166.0 \\
CVQZ version 0.1& 2.2353 & 1003.5  & 14.73 & 165.0 \\
CVQZ version 0.2& 2.2419 &  991.7  & 13.93 & 166.2 \\
Final CVQZ (incl. +2p)& 2.2479 &  979.0  & 14.32 & 167.8 \\
CV5Z version 0.1& 2.2406 &  989.8  & 15.31 & 168.8 \\
CV5Z version 0.2& 2.2445 &  985.3  & 15.81 & 169.4 \\
Final CV5Z (incl. +2p)& 2.2461 &  982.6  & 15.84 & 170.0 \\
Expt.$^a$& 2.240164(10) & 986.6484(41) & 15.54 & 170.972$^b$ \\
\hline
\multicolumn{5}{c}{K$_2$}\\
\hline
CVDZ Feller     & 4.1250 &   77.1  &  0.45 &  27.5 \\ 
CVDZ version 0.1& 4.1507 &   74.6  &  0.42 &  27.2 \\
CVDZ version 0.2& 4.2223 &   69.9  &  0.43 &  24.5 \\
Final CVDZ (incl. +2p)& 4.0511 &   88.5  &  0.29 &  50.0 \\
CVTZ Feller     & 3.9956 &   87.2  &  0.40 &  36.9 \\
CVTZ version 0.1& 3.9561 &   92.1  &  0.32 &  43.4 \\
CVTZ version 0.2& 4.0180 &   78.1  &  0.38 &  30.9 \\
Final CVTZ (incl. +2p)& 3.9837 &   90.1  &  0.25 &  50.2 \\
CVQZ Feller     & 3.9434 &   91.8  &  0.34 &  41.7 \\
CVQZ version 0.1& 3.9522 &   93.0  &  0.29 &  48.7 \\
CVQZ version 0.2& 3.9502 &   90.9  &  0.35 &  44.0 \\
Final CVQZ (incl. +2p)& 3.9557 &   92.0  &  0.33 &  51.2 \\
CV5Z version 0.1& 3.9440 &   92.9  &  0.29 &  51.6 \\
Final CV5Z (incl. +2p)& 3.9380 &   93.0x &  0.34 &  52.2 \\
Expt.$^c$& 3.92435(2) & 92.39766(47) & 0.32485(12) & 53.243(2) \\
\hline\hline
\end{tabular}

\begin{flushleft}
(a) \auth{H.}{Uehara} \auth{K.}{Horiai} \andauth{T.}{Konno} \JMST{413}{457}{1997}. \\

(b) Ref.\cite{CRC81}, corrected to 0 K using spectroscopic constants given to the left. \\

(c) \auth{C.}{Amiot} \auth{J.}{Verg\`es} \andauth{C. E.}{Fellows} \JCP{103}{3350}{1995}.
\end{flushleft}
\end{table}
\endgroup

\begingroup
\squeezetable
\begin{table}[h]
\caption{Basis set convergence of ionization potentials of K and Ca (eV),
with and without low-exponent $p$ functions added.\label{tab:IPs}}
\begin{tabular}{l|r|r|r|r}
\hline\hline
\multicolumn{5}{c}{K: IP(expt.)=4.34077$^a$} \\
\hline
 & \multicolumn{2}{c}{CCSD(T)} & \multicolumn{2}{c}{DK-CCSD(T)}\\
 &    & + 2p &   & +2p \\
\hline
CVDZ              &  4.068  &  4.168  &  4.086  & 4.185 \\
CVTZ              &  4.218  &  4.275  &  4.236  & 4.292 \\ 
CVQZ              &  4.300  &  4.301  &  4.317  & 4.318 \\
CV5Z              &  4.308  &  4.308  &  4.326  & 4.326 \\
CV$\infty$Z       &  4.316  &  4.315  &  4.335  & 4.334 \\
\hline
\multicolumn{5}{c}{Ca: IP(expt.)=6.11316$^a$} \\
\hline
 & \multicolumn{2}{c}{CCSD(T)} & \multicolumn{2}{c}{DK-CCSD(T)}\\  
 &    & + 1p &   & +1p \\
\hline
CVDZ              &  5.598  & 6.003   &  5.619  & 6.022 \\
CVTZ              &  5.797  & 6.050   &  5.818  & 6.069 \\
CVQZ              &  6.038  & 6.075   &  6.085  & 6.094 \\
CV5Z              &  6.073  & 6.083   &  6.094  & 6.102 \\
CV$\infty$Z       &  6.110  & 6.091   & 6.103   & 6.110 \\
\hline\hline
\end{tabular}

Basis set limits using expression\cite{Hal98}: $E_\infty\approx E_L+(E_L-E_{L-1})/((L/L-1)^3-1)$\\

(a) {\it Handbook of Chemistry and Physics} (CRC Press, Boca Raton, FL, 2001), p. 10-175.

\end{table}
\endgroup

\begin{table}[h]

\caption{Valence correlation, core-valence and deep-core correlation exponents in the
CVTZ and CCVTZ basis sets for potassium\label{tab:exponents}.}

\begin{tabular}{l|c|c|c}
\hline\hline
&Valence & Core & Deep core \\
\hline
s & 0.0181, 0.0501 & 0.346, 1.779 & 4.281, 20.055 \\
p & 0.0275, 0.0598 & 0.529, 2.599 & 7.240, 16.761 \\
d & 0.0599, 0.431 & 1.036, 2.586 & 9.724, 28.720 \\
f & 0.0888 & 1.129 & 15.150 \\
\hline\hline
\end{tabular}

\end{table}
\newpage
\begingroup
\squeezetable
\begin{table}[h]
\caption{Contracted basis set sizes and total numbers of basis functions for the CV$n$Z basis sets for group 1 and 2 elements\label{tab:contract}}
\begin{tabular}{l|lr|lr|lr|lr}
\hline\hline
      &\multicolumn{2}{c}{CVDZ} &\multicolumn{2}{c}{CVTZ}&\multicolumn{2}{c}{CVQZ}&\multicolumn{2}{c}{CV5Z} \\
\hline
Li,Be & [4s3p1d] & 18 & [6s5p3d1f] & 43 &  [8s7p5d3f1g] &  84 & [10s9p7d5f3g1h]  & 145 \\
Na,Mg & [5s4p2d] & 27 & [7s6p4d2f] & 59 &  [9s8p6d4f2g] & 109 & [11s10p8d6f4g2h] & 181 \\
K,Ca  & [6s5p2d] & 31 & [8s7p4d2f] & 63 & [10s9p6d4f2g] & 113 & [12s11p8d6f4g2h] & 185 \\
\hline
\end{tabular}
\end{table}

\endgroup
\newpage
\begingroup
\squeezetable
\begin{table}[h]
\caption{Computed and observed spectroscopic constants (kJ/mol, \AA, cm$^{-1}$ as appropriate) 
for alkali and alkaline earth metal hydrides\label{tab:hydrides}.}
\begin{tabular}{l|rr|rr|rr|rr}
\hline\hline
      & \multicolumn{2}{c}{$r_e$} & \multicolumn{2}{c}{$\omega_e$} & \multicolumn{2}{c}{$\omega_ex_e$} & \multicolumn{2}{c}{$D_0$} \\
      & Val & RIV & Val & RIV & Val & RIV & Val & RIV \\
\hline
\multicolumn{9}{c}{LiH}\\
\hline
CVDZ     & 1.6191 &  1.6106 & 1369.1  & 1379.6 & 20.80 & 21.55 & 209.1 & 210.6 \\
CVTZ     & 1.6081 &  1.5994 & 1394.9  & 1400.1 & 22.72 & 22.08 & 227.6 & 228.7 \\
CVQZ     & 1.6069 &  1.5960 & 1393.1  & 1405.3 & 23.61 & 22.84 & 231.4 & 232.6 \\
CV5Z     & 1.6074 &  1.5954 & 1391.8  & 1405.7 & 22.78 & 23.21 & 232.4 & 233.6 \\
Expt.$^a$& \multicolumn{2}{c}{1.5955991(1)} & \multicolumn{2}{c}{1405.49805(76)} &
\multicolumn{2}{c}{23.167899(714)} & \multicolumn{2}{c}{244.33, 234.354(4)$^g$} \\
\hline
\multicolumn{9}{c}{BeH}\\
\hline
CVDZ     & 1.3587 &  1.3556 & 2036.1  & 2033.9 & 34.35 & 34.26 & 177.4 & 178.1 \\
CVTZ     & 1.3498 &  1.3448 & 2039.9  & 2053.3 & 35.40 & 35.79 & 191.3 & 193.1 \\
CVQZ     & 1.3458 &  1.3411 & 2054.9  & 2065.6 & 36.32 & 36.65 & 195.9 & 197.7 \\
CV5Z     & 1.3456 &  1.3407 & 2055.3  & 2065.9 & 36.57 & 36.68 & 196.8 & 198.7 \\
Expt.$^b$& \multicolumn{2}{c}{1.3411(1)} & \multicolumn{2}{c}{2061.66} &
\multicolumn{2}{c}{37.18} & \multicolumn{2}{c}{196.2, $193.3 \pm 1.3$$^g$} \\
\hline
\multicolumn{9}{c}{NaH}\\
\hline
CVDZ     & 1.9237 &  1.9078 & 1116.3  & 1115.7 & 17.41 & 17.58 & 158.0 & 158.7 \\
CVTZ     & 1.9232 &  1.8949 & 1131.6  & 1154.0 & 16.46 & 18.56 & 176.1 & 176.5 \\
CVQZ     & 1.9256 &  1.8907 & 1122.8  & 1159.3 & 10.10 & 17.89 & 179.7 & 180.2 \\
CV5Z     & 1.9235 &  1.8883 & 1137.8  & 1171.1 & 18.43 & 19.67 & 181.7 & 181.2 \\
Expt.$^c$& \multicolumn{2}{c}{1.88703} & \multicolumn{2}{c}{$1171.968 \pm 0.012$} &
\multicolumn{2}{c}{$19.703 \pm 0.010$} & \multicolumn{2}{c}{(181), $182.03\pm0.25$$^g$} \\
\hline
\multicolumn{9}{c}{MgH}\\
\hline
CVDZ     & 1.7442 &  1.7398 & 1468.1  & 1458.9 & 27.22 & 26.31 & 107.6 & 106.3 \\
CVTZ     & 1.7434 &  1.7332 & 1496.6  & 1499.7 & 28.36 & 28.59 & 123.3 & 120.3 \\
CVQZ     & 1.7413 &  1.7299 & 1494.5  & 1498.1 & 28.46 & 29.10 & 127.1 & 122.9 \\
CV5Z     & 1.7401 &  1.7289 & 1493.2  & 1498.8 & 28.25 & 29.31 & 128.3 & 123.5 \\
Expt.$^d$& \multicolumn{2}{c}{1.729828(2)} & \multicolumn{2}{c}{1495.2632(22)} &
\multicolumn{2}{c}{31.64139(84)} & \multicolumn{2}{c}{129, $122.7\pm2.9$$^g$} \\
\hline
\multicolumn{9}{c}{KH}\\
\hline
CVDZ     & 2.3567 &  2.3232 &  907.6  &  902.1 & 12.58 & 11.42 & 141.0 & 136.1 \\
CVTZ     & 2.3230 &  2.2591 &  950.7  &  976.3 & 14.97 & 15.41 & 167.5 & 161.2 \\
CVQZ     & 2.3241 &  2.2479 &  944.4  &  979.0 & 14.41 & 14.32 & 173.3 & 167.8 \\
DK-CVQZ  &        &  2.2447 &         &  981.7 &       & 14.38 &       & 167.4 \\
CV5Z     & 2.3274 &  2.2461 &  938.8  &  982.6 & 14.81 & 15.84 & 174.8 & 170.0 \\
Expt.$^e$& \multicolumn{2}{c}{2.240164(10)} & \multicolumn{2}{c}{986.6484(41)} &
\multicolumn{2}{c}{15.54} & \multicolumn{2}{c}{179, 170.972$^g$} \\
\hline
\end{tabular}

\begin{center}
(continued on next page)
\end{center}
\end{table}
\endgroup
\clearpage
\begingroup
\addtocounter{table}{-1}
\squeezetable
\begin{table}
\caption{(continued)}
\begin{tabular}{l|rr|rr|rr|rr}
\hline
\multicolumn{9}{c}{CaH}\\
\hline
CVDZ     & 2.1453 &  2.0941 & 1231.5  & 1227.4 & 20.04 & 18.17 & 114.3 & 115.7 \\
CV(D+d)Z &        &  2.0539 &         & 1249.6 &       & 18.21 &       & 121.8 \\
CV(D+2d)Z&        &  2.0323 &         & 1260.3 &       & 18.25 &       & 126.7 \\
CV(D+3d)Z&        &  2.0227 &         & 1290.5 &       & 20.56 &       & 129.8 \\
CVTZ     & 2.1341 &  2.0415 & 1253.2  & 1273.9 & 18.70 & 17.29 & 141.5 & 148.4 \\
CV(T+d)Z &        &  2.0204 &         & 1279.1 &       & 16.85 &       & 153.7 \\
CV(T+2d)Z&        &  2.0149 &         & 1283.8 &       & 17.46 &       & 154.7 \\
CVQZ     & 2.0899 &  2.0063 & 1270.4  & 1297.9 & 20.62 & 18.80 & 158.5 & 162.5 \\
CV(Q+d)Z &        &  2.0048 &         & 1298.7 &       & 18.72 &       & 162.9 \\
DK-CVQZ  &        &  2.0077 &         & 1296.9 &       & 18.53 &       & 160.4 \\
DK-CV(Q+d)Z&      &  2.0062 &         & 1297.9 &       & 18.65 &       & 160.8 \\
CV5Z     & 2.0536 &  2.0027 & 1271.9  & 1299.1 & 18.68 & 18.93 & 169.3 & 165.2 \\
Expt.$^f$& \multicolumn{2}{c}{2.002366(16)} & \multicolumn{2}{c}{1298.3999(40)} &
\multicolumn{2}{c}{19.1842(28)} & \multicolumn{2}{c}{$\leq164$, 164.1$^g$} \\
\hline\hline
\end{tabular}

\begin{flushleft}
\noindent (a) (LiH)  \auth{M.}{Dulick} \auth{K.-Q.}{Zhang} \auth{B.}{Guo} \andauth{P.F.}{Bernath} \JMSP{188}{14}{1998}. \\
\noindent (b) (BeH) (1)  \twoauth{R.}{Colin}{D.}{De Greef} \CJP{53}{2142}{1975}. (2)  \auth{J. M. L.}{Martin} \CPL{283}{283}{1998}. \\
\noindent (c) (NaH)  \auth{F. P.}{Pesl} \auth{S.}{Lutz}  \andauth{K.}{Bergmann} \jcite{Eur. Phys. J. D.}{10}{247}{2000}. \\
\noindent (d) (MgH)  \auth{B.}{Lemoine} \auth{C.}{Demuynck} \auth{J.L.}{Destombes} \andauth{P.B.}{Davies} \JCP{89}{673}{1988}. \\
\noindent (e) (KH) \auth{H.}{Uehara} \auth{K.}{Horiai} \andauth{T.}{Konno} \JMST{413}{457}{1997}. \\
\noindent (f) (CaH)  \auth{D.}{Petitprez}  \auth{B.}{Lemoine}  \auth{C.}{Demuynck}  \auth{J.L.}{Destombes}  \andauth{B.}{Macke} \JCP{91}{4462}{1989}. \\
\noindent (g) First value from Huber and Herzberg\cite{Hub79}, second from CRC Handbook\cite{CRC81} corrected to 0 K with expt. $\omega_ex_e$ and $\omega_e$. \\
\end{flushleft}

\end{table}
\endgroup

\begingroup
\squeezetable
\begin{table}[h]
\caption{Computed and observed spectroscopic constants (kJ/mol, \AA, cm$^{-1}$ as appropriate)
for alkali and alkaline earth metal diatomics\label{tab:dimetals}.}
\begin{tabular}{l|rr|rr|rr|rr}
\hline\hline
      & \multicolumn{2}{c}{$r_e$} & \multicolumn{2}{c}{$\omega_e$} & \multicolumn{2}{c}{$\omega_ex_e$} & \multicolumn{2}{c}{$D_0$} \\
      & Val & RIV & Val & RIV & Val & RIV & Val & RIV \\
\hline
\multicolumn{9}{c}{Li$_2$}\\
\hline
CVDZ     & 2.7284 &  2.6963 &  340.4  &  346.7 &  2.32 &  2.44 &  92.0 &  95.5 \\
CVTZ     & 2.7006 &  2.6802 &  344.7  &  348.8 &  2.50 &  2.53 &  97.3 &  98.2 \\
CVQZ     & 2.6984 &  2.6756 &  346.6  &  351.0 &  2.53 &  2.58 &  98.5 &  99.2 \\
CV5Z     & 2.6985 &  2.6741 &  346.6  &  351.3 &  2.54 &  2.59 &  98.8 &  99.4 \\
Expt.$^a$& \multicolumn{2}{c}{2.6729} & \multicolumn{2}{c}{351.43} &
\multicolumn{2}{c}{2.610} & \multicolumn{2}{c}{100.9, $107.45\pm4$$^h$} \\
\hline
\multicolumn{9}{c}{LiNa}\\
\hline
CVDZ     & 2.9679 &  2.9386 &  244.1  &  247.5 &  1.38 &  1.45 &  77.0 &  79.9 \\
CVTZ     & 2.9505 &  2.9051 &  247.6  &  253.5 &  1.50 &  1.55 &  81.0 &  82.5 \\
CVQZ     & 2.9496 &  2.8953 &  248.6  &  256.1 &  1.56 &  1.64 &  81.7 &  83.0 \\
CV5Z     & 2.9500 &  2.8910 &  248.9  &  256.4 &  1.55 &  1.61 &  82.0 &  83.1 \\
Expt.$^b$& \multicolumn{2}{c}{2.88881(2)} & \multicolumn{2}{c}{256.4577(14)} &
\multicolumn{2}{c}{1.5808(6)} & \multicolumn{2}{c}{N/A, $84.728\pm0.001$$^h$} \\
\hline
\multicolumn{9}{c}{Na$_2$}\\
\hline
CVDZ     & 3.2048 &  3.1829 &  149.9  &  152.8 &  0.80 &  1.48 &  65.0 &  67.9 \\
CVTZ     & 3.1779 &  3.1007 &  151.7  &  158.4 &  0.66 &  0.69 &  68.4 &  70.9 \\
CVQZ     & 3.1773 &  3.0884 &  151.7  &  158.7 &  0.69 &  0.74 &  68.8 &  70.6 \\
DK-CVQZ  &        &  3.0819 &         &  159.4 &       &  0.77 &       &  70.9 \\
CV5Z     & 3.1783 &  3.0822 &  152.1  &  159.3 &  0.67 &  0.71 &  69.1 &  70.7 \\
Expt.$^c$& \multicolumn{2}{c}{3.0795(1)} & \multicolumn{2}{c}{159.103(3)} &
\multicolumn{2}{c}{0.7190(6)} & \multicolumn{2}{c}{69, $71.0173\pm0.0001$$^h$} \\
\hline
\multicolumn{9}{c}{NaK}\\
\hline
CVDZ     & 3.6793 &  3.6072 &  115.7  &  118.4 &  0.45 &  0.44 &  58.0 &  59.6 \\
CVTZ     & 3.6666 &  3.5379 &  116.6  &  121.3 &  0.43 &  0.44 &  61.3 &  60.9 \\
CVQZ     & 3.6671 &  3.5185 &  116.7  &  122.8 &  0.44 &  0.44 &  61.8 &  61.4 \\
CV5Z     & 3.6661 &  3.5070 &  117.0  &  124.0 &  0.44 &  0.44 &  61.9 &  61.9 \\
Expt.$^d$& \multicolumn{2}{c}{3.49903} & \multicolumn{2}{c}{123.993} &
\multicolumn{2}{c}{0.3045} & \multicolumn{2}{c}{60, $64.089\pm0.008$$^h$} \\
\hline
\multicolumn{9}{c}{Mg$_2$}\\
\hline
CVDZ     & 4.9450 &  4.8450 &   17.1  &   18.8 &  0.60 &  0.62 &   1.0 &   1.3 \\
CVTZ     & 4.0581 &  4.0333 &   41.0  &   43.6 &  1.65 &  1.51 &   3.0 &   3.3 \\
CVQZ     & 3.9748 &  3.9717 &   46.6  &   46.0 &  1.68 &  1.64 &   3.9 &   3.8 \\
CV5Z     & 3.9579 &  3.9709 &   47.6  &   45.9 &  1.67 &  1.50 &   4.1 &   3.9 \\
Expt.$^e$& \multicolumn{2}{c}{3.8905} & \multicolumn{2}{c}{51.121} &
\multicolumn{2}{c}{1.645} & \multicolumn{2}{c}{4.83, $7.037\pm0.004$$^h$} \\
\hline
\multicolumn{9}{c}{K$_2$}\\
\hline   
CVDZ     & 4.1626 &  4.0511 &   85.4  &   88.5 &  0.28 &  0.29 &  48.7 &  50.0 \\
CVTZ$^f$ & 4.1684 &  3.9837 &   85.8  &   90.1 &  0.27 &  0.25 &  51.1 &  50.2 \\
CVQZ     & 4.1694 &  3.9557 &   85.8  &   92.0 &  0.28 &  0.33 &  51.6 &  51.2 \\
DK-CVQZ  &        &  3.9407 &         &   92.9 &       &  0.21 &       &  52.1 \\
CV5Z     & 4.1693 &  3.9380 &   85.6  &   93.0 &  0.28 &  0.34 &  51.7 &  52.2 \\
Expt.$^g$& \multicolumn{2}{c}{3.92435(2)} & \multicolumn{2}{c}{92.39766(47)} &
\multicolumn{2}{c}{0.32485(12)} & \multicolumn{2}{c}{49.6, 52.88(2)$^h$, 53.243(2)$^g$} \\
\hline
\end{tabular}

\vspace*{0.25in}
(continued on next page)
\end{table}
\endgroup

\newpage

\begingroup
\squeezetable
\addtocounter{table}{-1}
\begin{table}
\caption{(continued)}
\begin{tabular}{l|rr|rr|rr|rr}
\hline
      & \multicolumn{2}{c}{$r_e$} & \multicolumn{2}{c}{$\omega_e$} & \multicolumn{2}{c}{$\omega_ex_e$} & \multicolumn{2}{c}{$D_0$} \\
      & Val & RIV & Val & RIV & Val & RIV & Val & RIV \\
\hline
\multicolumn{9}{c}{Ca$_2$}\\
\hline   
CVDZ     & 5.0096 &  4.7730 &   26.8  &   33.8 &  1.11 &  1.23 &   2.5 &   3.4 \\
CV(D+d)Z &        &  4.7956 &         &   32.5 &       &  1.17 &       &   3.5 \\
CV(D+2d)Z&        &  4.7642 &         &   33.5 &       &  1.19 &       &   3.7 \\
CV(D+3d)Z&        &  4.7554 &         &   33.9 &       &  1.17 &       &   3.8 \\
CVTZ     & 4.5409 &  4.4185 &   53.6  &   56.1 &  1.05 &  1.07 &   8.3 &   9.0 \\
CV(T+d)Z &        &  4.3932 &         &   57.2 &       &  1.08 &       &   9.6 \\
CV(T+2d)Z&        &  4.3899 &         &   57.2 &       &  1.09 &       &   9.6 \\
CVQZ     & 4.4385 &  4.3440 &   60.6  &   60.9 &  1.05 &  1.07 &  10.7 &  11.0 \\
CV(Q+d)Z &        &  4.3417 &         &   61.0 &       &  1.07 &       &  10.9 \\
DK-CVQZ  &        &  4.3313 &         &   59.1 &       &  0.99 &       &  11.0 \\
DK-CV(Q+d)Z&      &  4.3289 &         &   59.2 &       &  1.00 &       &  11.1 \\
CV5Z     & 4.4005 &  4.3261 &   63.0  &   61.5 &  1.08 &  1.07 &  12.1 &  11.5 \\
Expt.$^i$& \multicolumn{2}{c}{4.2773, 4.277} & \multicolumn{2}{c}{64.93, 65.07} &
\multicolumn{2}{c}{1.065, 1.09} & \multicolumn{2}{c}{12.4$\pm$1, 13.1} \\
\hline\hline
\end{tabular}

\begin{flushleft}
\noindent (a) (Li$_2$) Huber and Herzberg\cite{Hub79}. \\
\noindent (b) (LiNa)  \auth{C.E.}{Fellows} \JPC{94}{5855}{1991}. \\
\noindent (c) (Na$_2$)  \twoauth{O.}{Babaky}{K. J.}{Hussein} \CJP{67}{912}{1989}. \\
\noindent (d) (NaK)  \twoauth{A.}{Krou-Adohi}{S.}{Giraud-Cotton} \JMSP{190}{171}{1998}. \\
\noindent (e) (Mg$_2$) Huber and Herzberg\cite{Hub79}. \\
\noindent (f) (K$_2$) With CCVTZ deep-core correlation basis set and correlating all but $(1s)$ electrons
(RIV values given in parentheses): $r_e$=3.9809(3.9827) \AA, $\omega_e$=90.2(90.2) cm$^{-1}$, 
$\omega_ex_e$=0.25(0.25) cm$^{-1}$, $D_0$=50.3(50.2) kJ/mol.
\noindent (g) (K$_2$)  \auth{C.}{Amiot}  \auth{J.}{Verg\`es} \andauth{C.E.}{Fellows} \JCP{103}{3350}{1995}. \\
\noindent (h) First value from Huber and Herzberg\cite{Hub79}, second from CRC Handbook\cite{CRC81} corrected to 0 K with expt. $\omega_ex_e$ and $\omega_e$. \\
\noindent (i) First set of values: Huber and Herzberg\cite{Hub79}, from \twoauth{W. J.}{Balfour}{R. F.}{Whitlock} \CJP{53}{472}{1975}; Second set of values: \auth{C. R.}{Vidal} \JCP{72}{1864}{1980}, 
\twoauth{V. E.}{Bondybey}{J. H.}{English} \CPL{111}{195}{1984}.\\
\end{flushleft}


\end{table}
\endgroup

\newpage

\begingroup
\squeezetable

\begin{table}[h]
\caption{Computed and observed spectroscopic constants (kJ/mol, \AA, cm$^{-1}$ as appropriate)
for metal halides\label{tab:halides}.}
\begin{tabular}{l|rrr|rrr|rrr|rrr}
\hline\hline
      & \multicolumn{3}{c}{$r_e$} & \multicolumn{3}{c}{$\omega_e$} & \multicolumn{3}{c}{$\omega_ex_e$} & \multicolumn{3}{c}{$D_0$} \\
      & Val & RIV & All &Val & RIV & All & Val & RIV & All & Val & RIV & All \\
\hline
\multicolumn{13}{c}{LiF}\\
\hline
CVDZ     & 1.6074 & 1.5983 & 1.5977 &  870.6 &  876.1 &  876.9 &  8.16 &  8.18 &  8.18 & 539.7 & 543.8 & 544.8 \\
CVTZ     & 1.5903 & 1.5747 & 1.5741 &  885.5 &  891.2 &  891.5 &  7.79 &  7.70 &  7.74 & 558.7 & 562.4 & 563.0 \\
CVQZ     & 1.5817 & 1.5671 & 1.5665 &  894.4 &  906.1 &  906.6 &  8.00 &  8.12 &  8.13 & 568.9 & 572.7 & 573.1 \\
CV5Z     & 1.5808 & 1.5656 & 1.5650 &  895.9 &  908.0 &  908.6 &  8.04 &  8.12 &  8.14 & 571.0 & 575.0 & 575.6 \\
Expt.$^a$& \multicolumn{3}{c}{1.5638648(3)} & \multicolumn{3}{c}{910.57272(10)} &
\multicolumn{3}{c}{8.207956(46)} & \multicolumn{3}{c}{570, $537\pm21$$^m$} \\
\hline
\multicolumn{13}{c}{LiCl}\\
\hline
CVDZ     & 2.0900 & 2.0819 & 2.0806 &  600.9 &  607.4 &  608.7 &  3.57 &  3.97 &  3.95 & 434.5 & 436.3 & 437.4 \\
CVTZ     & 2.0486 & 2.0375 & 2.0351 &  626.9 &  633.1 &  634.3 &  4.26 &  4.33 &  4.34 & 454.3 & 455.9 & 456.5 \\
CVQZ     & 2.0430 & 2.0291 & 2.0266 &  632.2 &  640.1 &  641.0 &  4.38 &  4.45 &  4.45 & 465.5 & 467.4 & 468.0 \\
Expt.$^b$& \multicolumn{3}{c}{2.0206719(2)} & \multicolumn{3}{c}{642.95821(14)} &
\multicolumn{3}{c}{4.475085(57)} & \multicolumn{3}{c}{467, $466\pm13$$^m$} \\
\hline
\multicolumn{13}{c}{BeF}\\
\hline
CVDZ     & 1.4148 & 1.4083 & 1.4078 & 1190.9 & 1197.6 & 1198.9 &  8.27 &  8.52 &  8.54 & 509.9 & 513.4 & 514.6 \\
CVTZ     & 1.3741 & 1.3696 & 1.3690 & 1240.1 & 1245.3 & 1246.4 &  9.13 &  9.17 &  9.19 & 546.2 & 549.2 & 549.9 \\
CVQZ     & 1.3689 & 1.3635 & 1.3531 & 1254.8 & 1264.1 & 1264.3 &  9.30 &  9.39 &  9.40 & 558.8 & 562.8 & 563.1 \\
CV5Z     & 1.3678 & 1.3623 & 1.3617 & 1255.3 & 1264.5 & 1265.4 &  9.27 &  9.34 &  9.35 & 561.0 & 565.1 & 565.8 \\
Expt.$^c$& \multicolumn{3}{c}{1.36075(3)} & \multicolumn{3}{c}{1265.54(10)} &
\multicolumn{3}{c}{9.422(28)} & \multicolumn{3}{c}{564, 604, $573\pm42$$^m$} \\
\hline
\multicolumn{13}{c}{BeCl}\\
\hline
CVDZ     & 1.8323 & 1.8269 & 1.8251 &  817.4 &  816.5 &  817.8 &  5.12 &  4.89 &  4.89 & 341.3 & 343.3 & 344.7 \\
CVTZ     & 1.8165 & 1.8117 & 1.8087 &  833.1 &  837.6 &  839.6 &  4.87 &  4.90 &  4.93 & 361.5 & 363.2 & 364.8 \\
CVQZ     & 1.8076 & 1.8018 & 1.7989 &  839.3 &  844.5 &  846.5 &  4.91 &  4.94 &  4.94 & 373.4 & 375.9 & 377.4 \\
Expt.$^d$& \multicolumn{3}{c}{1.7971} & \multicolumn{3}{c}{846.7} &
\multicolumn{3}{c}{4.85(3)} & \multicolumn{3}{c}{435, 385, 333, 384.8$\pm9.2^m$} \\
\hline
\multicolumn{13}{c}{NaF}\\
\hline
CVDZ     & 1.9869 & 1.9485 & 1.9478 &  528.2 &  527.2 &  527.7 &  4.59 &  4.03 &  4.03 & 449.4 & 455.4 & 456.3 \\
CVTZ     & 1.9966 & 1.9412 & 1.9402 &  532.1 &  523.9 &  524.4 &  4.34 &  3.46 &  3.49 & 462.1 & 466.1 & 466.7 \\
CVQZ     & 1.9967 & 1.9318 & 1.9311 &  534.0 &  530.4 &  530.6 &  4.35 &  3.54 &  3.54 & 469.0 & 473.2 & 473.6 \\
CV5Z     & 1.9916 & 1.9291 & 1.9284 &  537.8 &  533.8 &  534.0 &  4.45 &  3.53 &  3.53 & 473.1 & 475.4 & 475.9 \\
Expt.$^e$& \multicolumn{3}{c}{1.9259455(2)} & \multicolumn{3}{c}{535.65805(21)} &
\multicolumn{3}{c}{3.57523(13)} & \multicolumn{3}{c}{(514), 516$^m$} \\
\hline
\multicolumn{13}{c}{NaCl}\\
\hline
CVDZ     & 2.4278 & 2.4117 & 2.4099 &  342.8 &  345.9 &  346.8 &  1.76 &  1.79 &  1.80 & 383.9 & 384.4 & 385.3 \\
CVTZ     & 2.4133 & 2.3868 & 2.3845 &  346.5 &  354.2 &  354.4 &  1.62 &  1.73 &  1.73 & 396.2 & 394.2 & 394.6 \\
CVQZ     & 2.4064 & 2.3729 & 2.3705 &  350.6 &  360.2 &  360.7 &  1.71 &  1.71 &  1.71 & 407.5 & 404.8 & 405.2 \\
Expt.$^f$& \multicolumn{3}{c}{2.3607941(4)} & \multicolumn{3}{c}{364.684163(391)} &
\multicolumn{3}{c}{1.776085(189)} & \multicolumn{3}{c}{408, $409.3\pm8$$^m$} \\
 \hline
\multicolumn{13}{c}{MgF}\\
\hline
CVDZ     & 1.7854 & 1.7782 & 1.7775 &  697.9 &  698.9 &  699.6 &  4.19 &  4.18 &  4.18 & 418.2 & 418.6 & 419.7 \\
CVTZ     & 1.7772 & 1.7627 & 1.7622 &  701.0 &  704.8 &  704.4 &  4.00 &  3.96 &  3.94 & 433.3 & 431.4 & 431.9 \\
CVQZ     & 1.7700 & 1.7541 & 1.7536 &  708.4 &  716.4 &  716.8 &  4.07 &  4.20 &  4.21 & 443.4 & 440.6 & 441.0 \\
CV5Z     & 1.7659 &        &        &  711.7 &        &        &  4.18 &       &       & 447.1 &       &       \\
Expt.$^g$& \multicolumn{3}{c}{1.7499371(1)} & \multicolumn{3}{c}{720.14042(30)} &
\multicolumn{3}{c}{4.26018(16)} & \multicolumn{3}{c}{458, $458.5\pm5.0$$^m$} \\
\hline
\end{tabular}

%
\end{table}
\endgroup

\clearpage 
\thispagestyle{empty}
\begingroup
\squeezetable
\addtocounter{table}{-1}

\begin{table}
\caption{(continued)}
%
%
 \begin{tabular}{l|rrr|rrr|rrr|rrr}
 \hline
       & \multicolumn{3}{c}{$r_e$} & \multicolumn{3}{c}{$\omega_e$} & \multicolumn{3}{c}{$\omega_ex_e$} & \multicolumn{3}{c}{$D_0$} \\
       & Val & RIV & All &Val & RIV & All & Val & RIV & All & Val & RIV & All \\
\hline
\multicolumn{13}{c}{MgCl}\\
\hline
CVDZ     & 2.2582 & 2.2558 & 2.2541 &  441.8 &  440.5 &  441.3 &  1.85 &  1.79 
&  1.79 & 290.5 & 289.3 & 290.4\\
CVTZ     & 2.2251 & 2.2137 & 2.2110 &  457.6 &  461.9 &  463.2 &  1.99 &  2.04 &  2.06 & 310.5 & 307.1 & 308.1 \\
CVQZ     & 2.2193 & 2.2053 & 2.2025 &  459.8 &  464.2 &  465.2 &  2.01 &  2.04 &  2.05 & 321.4 & 316.8 & 317.7 \\
Expt.$^h$& \multicolumn{3}{c}{2.196(1)} & \multicolumn{3}{c}{466.0(8)} &
\multicolumn{3}{c}{2.0(0)} & \multicolumn{3}{c}{317, $324.5\pm2.1$$^m$} \\
\hline
\multicolumn{13}{c}{KF}\\
\hline
CVDZ     & 2.2436 & 2.2514 & 2.2506 &  402.7 &  399.0 &  399.3 &  2.58 &  2.41 &  2.41 & 368.4 & 454.7 & 359.9 \\
CVTZ     & 2.1993 & 2.1980 & 2.1972 &  414.7 &  415.6 &  415.7 &  2.46 &  2.24 &  2.26 & 332.5 & 477.6 & 370.1 \\
CVQZ     & 2.1868 & 2.1800 & 2.1796 &  419.7 &  422.2 &  422.2 &  2.46 &  2.35 &  2.36 & 322.5 & 488.6 & 422.2 \\
CV5Z     &        &        &        &        &        &        &       &       &       & 319.6$^l$ & 494.5$^l$ & 435.6$^l$ \\
Expt.$^i$& \multicolumn{3}{c}{2.1714559(2)} & \multicolumn{3}{c}{426.261872(98)} &
\multicolumn{3}{c}{2.449801(44)} & \multicolumn{3}{c}{489, $494.5\pm2.5$$^m$} \\
\hline
\multicolumn{13}{c}{KCl}\\
\hline
CVDZ     & 2.9169 & 2.7709 & 2.7684 &  271.2 &  258.7 &  259.2 &  2.89 &  1.18 &  1.18 & 391.5 & 393.4 & 394.0 \\
CVTZ     & 2.9705 & 2.7045 & 2.7017 &  301.9 &  270.4 &  270.8 &  3.31 &  1.15 &  1.15 & 400.4 & 407.8 & 408.3 \\
CVQZ     & 2.9827 & 2.6829 & 2.6801 &  308.8 &  276.4 &  276.7 &  3.34 &  1.20 &  1.20 & 410.1 & 421.1 & 421.5 \\
Expt.$^f$& \multicolumn{3}{c}{2.666678(3)} & \multicolumn{3}{c}{280.07639(490)} &
\multicolumn{3}{c}{1.31330(338)} & \multicolumn{3}{c}{419, $430.5\pm8$$^m$} \\
\hline
\multicolumn{13}{c}{CaF}\\
\hline
CVDZ     & 2.1269 & 2.0406 & 2.0399 &  405.4 &  536.4 &  536.5 & 30.03 &  2.62 &  2.62 & 342.7 & 473.3 & 382.8 \\
CV(D+3d)Z&        & 1.9902 &        &        &  571.3 &        &       &  3.00 &       &       & 508.5 &       \\
CVTZ     & 2.1125 & 1.9889 & 1.9880 &  399.2 &  562.3 &  563.0 & 38.60 &  2.75 &  2.78 & 310.5 & 508.4 & 399.4 \\
CV(T+2d)Z&        & 1.9692 &        &        &  574.3 &        &       &  2.85 &       &       & 532.3 &       \\
CVQZ     & 1.9981 & 1.9597 & 1.9595 &  537.6 &  583.6 &  583.5 &  5.37 &  2.87 &  2.87 & 361.0 & 531.1 & 463.9 \\
CV(Q+d)Z &        & 1.9585 &        &        &  584.1 &        &       &  2.88 &       &       & 531.6 &       \\
Expt.$^j$& \multicolumn{3}{c}{1.9516403(1)} & \multicolumn{3}{c}{588.644(2)} &
\multicolumn{3}{c}{2.91194(6)} & \multicolumn{3}{c}{529, $524\pm21$$^m$} \\
\hline
\multicolumn{13}{c}{CaCl}\\
\hline
CVDZ     & 2.6598 & 2.5770 & 2.5752 &  324.2 &  335.4 &  336.1 &  1.22 &  1.16 &  1.15 &  24.8 & 352.8 & 353.7 \\
CV(D+3d)Z&        & 2.5049 &        &        &  348.2 &        &       &  1.27 &       &       & 368.4 &       \\
CVTZ     & 2.5969 & 2.4921 & 2.4891 &  340.9 &  353.5 &  354.1 &  1.27 &  1.27 &  1.27 & 370.0 & 382.0 & 382.8 \\
CV(T+2d)Z&        & 2.4704 &        &        &  357.1 &        &       &  1.27 &       &       & 388.5 &       \\
CVQZ     & 2.5402 & 2.4531 & 2.4501 &  351.5 &  366.0 &  366.5 &  1.34 &  1.35 &  1.35 & 396.7 & 404.1 & 404.9 \\
CV(Q+d)Z &        & 2.4519 &        &        &  366.4 &        &       &  1.35 &       &       & 404.5 &       \\
Expt.$^k$& \multicolumn{3}{c}{2.43674} & \multicolumn{3}{c}{370.201} &
\multicolumn{3}{c}{1.3732} & \multicolumn{3}{c}{395, $406\pm9$$^m$} \\
\hline\hline
\end{tabular}

\begin{flushleft}
\noindent (a) (LiF)  \auth{H.G.}{Hedderich}  \auth{F.R.}{Engleman Jr.}  \andauth{P.F.}{Bernath} \CJC{69}{1659}{1991}. \\
\noindent (b) (LiCl)  \auth{J.B.}{Burkholder}  \auth{P.D.}{Hammer}  \auth{C.J.}{Howard}  \auth{A.G.}{Maki}  \auth{G.}{Thompson}  \andauth{C.}{Chackerian Jr.} \JMSP{124}{139}{1987}. \\
\noindent (c) (BeF)  \twoauth{G.}{Tai}{R.D.}{Verma} \JMSP{173}{1}{1995}. \\
\noindent (d) (BeCl) Huber and Herzberg\cite{Hub79}. \\
\noindent (e) (NaF)  \auth{A.}{Muntianu} \auth{B.}{Guo} \andauth{P.F.}{Bernath} \JMSP{176}{274}{1996}. \\
\noindent (f) (NaCl, KCl)  \auth{R.S.}{Sam}  \auth{M.}{Dulick}  \auth{B.}{Guo}  \auth{K.-Q.}{Zhang}  \andauth{P.F.}{Bernath} \JMSP{183}{360}{1997}. \\
\noindent (g) (MgF)  \auth{B.E.}{Barber}  \auth{K.-Q.}{Zhang}  \auth{B.}{Guo}  \andauth{P.F.}{Bernath} \JMSP{169}{583}{1995}. \\
\noindent (h) (MgCl)  \auth{J.}{Rostas}  \auth{N.}{Shafizadeh}  \auth{G.}{Taieb}  \andauth{B.}{Bourguignon} \CP{142}{97}{1990}. \\
\noindent (i) (KF)  \auth{M.-C.}{Liu}  \auth{A.}{Muntianu}  \auth{K.-Q.}{Zhang}  \auth{P.}{Colarusso}  \andauth{P.F.}{Bernath} \JMSP{180}{188}{1996}. \\
\noindent (j) (CaF)  \auth{L.A.}{Kaledin}  \auth{J.C.}{Bloch}  \auth{M.C.}{McCarthy}  \andauth{R.W.}{Field} \JMSP{197}{289}{1999}. \\
\noindent (k) (CaCl)  \auth{L.-E.}{Berg}  \auth{L.}{Klynning}  \auth{H.}{Martin}  \auth{A.}{Pereira}  \andauth{P.}{Royen} \jcite{Phys. Scr.}{24}{23}{1981}. \\
\noindent (l) Single point energy calculation at experimental $r_e$. \\
\noindent (m) All except last values from Huber and Herzberg\cite{Hub79}, last value from CRC Handbook\cite{CRC81} corrected to 0 K with expt. $\omega_ex_e$ and $\omega_e$. \\
\end{flushleft}

\end{table}

\endgroup
\newpage

\begingroup
\squeezetable
\begin{table}[h]
\caption{Computed and observed spectroscopic constants (kJ/mol, \AA, cm$^{-1}$ as appropriate) 
for metal chalcogenides\label{tab:oxides}.}
\begin{tabular}{l|rrr|rrr|rrr|rrr}
\hline\hline
      & \multicolumn{3}{c}{$r_e$} & \multicolumn{3}{c}{$\omega_e$} & \multicolumn{3}{c}{$\omega_ex_e$} & \multicolumn{3}{c}{$D_0$} \\
      & Val & RIV & All &Val & RIV & All & Val & RIV & All & Val & RIV & All \\
\hline
\multicolumn{13}{c}{BeO}\\
\hline
CVDZ     & 1.3711 & 1.3674 & 1.3671 & 1366.3 & 1371.1 & 1372.0 & 11.47 & 11.22 & 11.18 & 355.0 & 359.2 & 360.5 \\
CVTZ     & 1.3456 & 1.3414 & 1.3407 & 1450.6 & 1457.1 & 1458.4 & 12.07 & 12.04 & 12.05 & 408.5 & 412.7 & 413.6 \\
CVQZ     & 1.3387 & 1.3334 & 1.3326 & 1470.7 & 1483.5 & 1484.2 & 12.07 & 12.11 & 12.12 & 425.5 & 431.3 & 431.9 \\
CV5Z     & 1.3373 & 1.3317 & 1.3308 & 1475.4 & 1488.2 & 1489.6 & 12.03 & 12.08 & 12.09 & 429.9 & 435.8 & 436.6 \\
Expt.$^a$& \multicolumn{3}{c}{1.3309} & \multicolumn{3}{c}{1487.32} & \multicolumn{3}{c}{11.830} & \multicolumn{3}{c}{444, $431.0\pm13.4$$^d$} \\
\hline
\multicolumn{13}{c}{BeS}\\
\hline
CVDZ     & 1.7764 & 1.7704 & 1.7722 &  960.2 &  966.2 &  964.8 &  5.74 &  5.81 &  5.81 & 260.1 & 264.2 & 263.4 \\
CVTZ     & 1.7607 & 1.7568 & 1.7551 &  977.5 &  982.6 &  983.7 &  5.82 &  5.83 &  5.86 & 290.4 & 293.0 & 293.8 \\
CVQZ     & 1.7519 & 1.7467 & 1.7443 &  990.9 &  996.6 &  998.3 &  5.87 &  5.88 &  5.90 & 306.2 & 309.4 & 310.5 \\
Expt.$^a$& \multicolumn{3}{c}{1.7415(3)} & \multicolumn{3}{c}{997.94} & \multicolumn{3}{c}{6.137} & {367, $368\pm59$$^d$} \\
\hline
\multicolumn{13}{c}{MgO}\\
\hline
CVDZ     & 1.7859 & 1.7763 & 1.7753 &  743.2 &  754.5 &  756.8 &  9.69 &  9.79 &  9.77 & 204.8 & 207.2 & 208.4 \\
CVTZ     & 1.7662 & 1.7500 & 1.7491 &  772.2 &  794.1 &  795.8 &  9.21 &  9.04 &  8.94 & 234.0 & 236.5 & 237.0 \\
CVQZ     & 1.7595 & 1.7421 & 1.7413 &  787.6 &  810.0 &  811.8 &  9.26 &  8.81 &  8.72 & 245.2 & 247.1 & 247.4 \\
Expt.$^b$& \multicolumn{3}{c}{1.7481722(9)} & \multicolumn{3}{c}{785.2183(6)} &
\multicolumn{3}{c}{5.1327(3)} & \multicolumn{3}{c}{(341), $359.7\pm12.6$$^d$} \\
\hline
\multicolumn{13}{c}{MgS}\\
\hline
CVDZ     & 2.1999 & 2.1952 & 2.1958 &  499.9 &  498.6 &  498.4 &  2.63 &  2.51 &  2.50 & 162.5 & 163.3 & 163.8 \\
CVTZ     & 2.1718 & 2.1600 & 2.1582 &  518.6 &  524.0 &  525.1 &  2.60 &  2.58 &  2.57 & 195.6 & 195.4 & 196.0 \\
CVQZ     & 2.1635 & 2.1491 & 2.1465 &  524.4 &  530.3 &  531.5 &  2.62 &  2.60 &  2.59 & 208.5 & 207.4 & 208.0 \\
Expt.$^a$& \multicolumn{3}{c}{2.1425} & \multicolumn{3}{c}{528.74} & \multicolumn{3}{c}{2.704} & \multicolumn{3}{c}{$\leq232$, 231$^d$} \\
\hline
\multicolumn{13}{c}{CaO}\\
\hline
CVDZ     & 2.0360 & 2.0362 & 2.0531 &  777.5 &  779.9 &  490.2 &  2.17 &  2.25 &  5.15 & 143.5 & 283.0 & 194.2 \\
CV(D+d)Z &        & 2.0395 &        &        &  557.1 &        &       &-10.94 &       &       & 295.0 &       \\
CV(D+2d)Z&        & 1.9382 &        &        &  477.8 &        &       &  7.97 &       &       & 308.0 &       \\
CV(D+3d)Z&        & 1.9095 &        &        &  565.5 &        &       & 11.78 &       &       & 318.8 &       \\
CVTZ     & 2.0206 & 1.9298 & 1.9284 &  824.3 &  523.1 &  525.7 & 31.47 &  3.37 &  3.74 &  63.0 & 332.5 & 230.3 \\
CV(T+d)Z &        & 1.8540 &        &        &  661.2 &        &       &  8.71 &       &       & 360.5 &       \\
CV(T+2d)Z&        & 1.8472 &        &        &  679.2 &        &       &  7.88 &       &       & 366.6 &       \\
CVQZ     & 2.0658 & 1.8308 & 1.8307 &  689.9 &  718.0 &  717.7 &  5.23 &  6.34 &  6.35 &  66.8 & 390.7 & 331.4 \\
CV(Q+d)Z &        & 1.8281 &        &        &  722.2 &        &       &  6.13 &       &       & 393.3 &       \\
CV(Q+2d)Z&        & 1.8274 &        &        &  723.3 &        &       &  6.11 &       &       & 393.9 &       \\
DK-CVQZ &         & 1.8333 &        &        &  709.6 &        &       & 6.73   &         &       &  385.0 &   \\
DK-CV(Q+d)Z &     & 1.8305 &        &        &  714.1 &        &       &  6.54 &       &       & 387.5 &       \\
CV5Z     &        &        &        &        &        &        &       &       &       &       & 406.4$^e$ &       \\
Expt.$^c$& \multicolumn{3}{c}{1.8222315(4)} & \multicolumn{3}{c}{732.01377(40)} & 
\multicolumn{3}{c}{4.81268(74)} & \multicolumn{3}{c}{$\geq459$, $398.7\pm16.7$$^d$} \\
\hline\hline
\end{tabular}

\begin{flushleft}
\noindent (a) (BeO,BeS,MgS) Huber and Herzberg\cite{Hub79}.\\
\noindent (b) (MgO) \auth{P.}{M\"urtz} \auth{H.}{Th\"ummel} \auth{C.}{Pfelzer} \andauth{W.}{Urban} \MP{86}{513}{1995}. \\
\noindent (c) (CaO)  \auth{C.}{Focsa}  \auth{A.}{Poclet}  \auth{B.}{Pinchemel}  \auth{R.J.}{Le Roy}  \andauth{P.F.}{Bernath} \JMSP{203}{330}{2000}. \\
\noindent (d) First value from Huber and Herzberg\cite{Hub79}, second from CRC Handbook\cite{CRC81} corrected to 0 K with expt. $\omega_ex_e$ and $\omega_e$. \\
\noindent (e) Single-point energy calculation at experimental $r_e$. Results with the CV(5+d)Z and CV(5+2d)Z basis sets
are 406.7 and 406.6 kJ/mol, respectively.\\
\end{flushleft}

\end{table}
\endgroup

\begingroup
\squeezetable
\begin{table}[h]
\caption{Computed $^2D\leftarrow{}^2S$ excitation energy (cm$^{-1}$) in Ca$^+$ as a function
of the basis set.\label{tab:Ca+}}
\begin{tabular}{lrrrr}
\hline\hline
           &   D      &    T         &   Q         &     5        \\
\hline
CVnZ       &38065      &    22546     &    13965    &     12374    \\
CV(n+d)Z   &28618      &    16539     &    12985    &     12898    \\
CV(n+2d)Z  &21528      &    14920     &    12851    &     12885    \\
CV(n+3d)Z  &19214      &    15155     &    ---      &     ---      \\
CV(Q+d)Z uncontracted& &              &    13240    &     ---      \\
DK-CV(Q+d)Z uncontracted& &           &    14201    &     ---      \\
Best estimate$^a$      & &           &             &     13846    \\
\hline\hline
\end{tabular}

\begin{flushleft}
All calculations at the CAS-PT3 level\cite{rs3} with a [Ne] core frozen. The experimental 
values are\cite{nist}: $^2D_{3/2}$~13650.19, $^2D_{5/2}$~13710.88, spin-orbit 
averaged 13686.60 cm$^{-1}$. 

(a) CV(5+2d)Z plus relativistic correction estimated as the difference between DK-CV(Q+d)Z and CV(Q+d)Z.
\end{flushleft}

\end{table}

\endgroup


\begin{thebibliography}{10}
\bibitem{stretch} By analogy with `even-tempered'\cite{Rue74} and `well-tempered'\cite{wtbs} basis sets, the
term `stretch-tuned' is borrowed from the musician's lingo. It refers
to the habit of piano tuners, in order to reduce strain
on the frame, to deviate from `even-tempered' tuning at the low and high ends
of the keyboard. Specifically, low notes are tuned flat, and high
notes sharp, relative to their even-tempered pitch.
\end{thebibliography}
\end{document}